# Weyl gauge-vector and complex dilaton scalar for conformal symmetry and its breaking

## Hans C. Ohanian[1]

**Abstract** Instead of the scalar "dilaton" field that is usually adopted to construct conformally invariant Lagrangians for gravitation, we here propose a hybrid construction, involving both a complex dilaton scalar and a Weyl gauge-vector, in accord with Weyl's original concept of a non-Riemannian conformal geometry with a transport law for length and time intervals, for which this gauge vector is required. Such a hybrid construction permits us to avoid the wrong sign of the dilaton kinetic term (the ghost problem) that afflicts the usual construction. The introduction of a Weyl gauge-vector and its interaction with the dilaton also has the collateral benefit of providing an explicit mechanism for spontaneous breaking of the conformal symmetry, whereby the dilaton and the Weyl gauge-vector acquire masses somewhat smaller than $m_P$ by the Coleman-Weinberg mechanism. Conformal symmetry breaking is assumed to precede inflation, which occurs later by a separate GUT or electroweak symmetry breaking, as in inflationary models based on the Higgs boson.

## 1 Introduction

Modifications of Einstein's gravitational theory that incorporate local conformal symmetry—that is, invariance under the transformation $g_{\mu\nu}(x) \to \mathrm{e}^{2\alpha(x)} g_{\mu\nu}(x)$, where $\alpha(x)$ is an arbitrary real function—have been exploited in attempts at the solution of various of theoretical problems, such as renormalization of the stress tensor, renormalization of quantum gravity, quantum mechanics of black holes, analytic solutions and geodesic completeness in the early universe, and the dynamics that lead to inflation by symmetry breaking.

The conventional Einstein theory, with the Lagrangian $(m_P{}^2/16\pi)\sqrt{-g}R$, lacks conformal symmetry. To endow this theory with a pedigree that includes conformal symmetry, we need to regard it as an effective field theory derived from a conformally symmetric precursor theory by spontaneous symmetry breaking. Such conformal symmetry breaking can arise in several ways: It can be collateral damage of the spontaneous breaking of the GUT and/or electroweak gauge symmetries, when the precursor zero-mass fermions and gauge bosons acquire masses and thereby spoil the conformal symmetry [1, 2, 3, 4, 5, 6]. Or else it can be implemented independently by a separate mechanism that directly breaks the conformal symmetry but leaves the GUT and/or electroweak gauge symmetries untouched [7, 8, 9, 10]. The conformal symmetry breaking may or may not be accompanied by inflation—if it is not, then inflation will have to occur later, in connection with the breaking of GUT or electroweak symmetry.

Conformal symmetry breaking by an independent mechanism is usually thought to arise from a scalar field $\phi(x)$, called the dilaton, which in the conformally symmetric regime has zero mass and a symmetric vacuum state but at lower temperatures and lower energies spontaneously settles into a nonsymmetric state, with a nonzero vacuum expectation value and a nonzero mass. The evolution of this dilaton scalar field toward its nonsymmetric state is governed by an effective potential, and there is an abundance (and overabundance) of models that achieve spontaneous symmetry breaking with the properties fancied by theorists.

―――――

[1] Department of Physics, University of Vermont, Burlington, VT 05405-0125, USA. hohanian@uvm.edu

The symmetry breaking that causes inflation is, likewise, thought to arise from a scalar field, called the inflaton. In principle, inflation could arise from breaking of the conformal symmetry (so inflaton = dilaton), and breaking of the GUT and electroweak symmetries could come later. However, in view of the experimental confirmation of the existence of the Higgs boson (with $m_H = 125$ GeV), it is tempting to identify this Higgs boson as the cause of inflation, as proposed in the neat model of Bezrukov and Shaposhnikov [11] in which inflation is treated as a consequence of electroweak spontaneous symmetry breaking and therefore occurs much later and separately from conformal symmetry breaking. According to this scenario, conformal symmetry breaking is not associated with inflation and does not affect the good agreement of the Bezrukov-Shaposhnikov model with the WMAP and Planck data [12].

As pointed out by Bars et al. [9], even if inflation is caused by the Higgs and is not coeval with conformal symmetry breaking, it is still desirable that the "full" theory should have as precursor a Lagrangian with conformal invariance, and this imposes restrictions on the form that the Lagrangian can take in the conformally broken regime. Bars et al. have contrived a general prescription for transforming ("lifting") Lagrangians that are not conformally invariant into precursor Lagrangians that are invariant. This provides a "conformalization" prescription for the selection of Lagrangians that are endowed with conformally invariant pedigrees (or "an underlying hidden conformal symmetry"), so with insertion of correction factors consisting of functions of the inflaton field $\phi(x)$ they can achieve conformal invariance. For instance, Bars et al. show how the Bezrukov-Shaposhnikov model emerges from a conformally invariant precursor Lagrangian, by breaking of the conformal symmetry.

Unfortunately, the construction of such precursor Lagrangians by insertion of a scalar field $\phi(x)$ faces a serious obstacle in that the appropriate kinetic term $\frac{1}{2} g^{\mu\nu} \partial_\mu \phi \partial_\nu \phi$ (which, with the signature $+---$ adopted here, contributes a positive energy density) must appear in the Lagrangian in the invariant combination

$$\pm \sqrt{-g} \left( \tfrac{1}{12} \phi^2 R - \tfrac{1}{2} g^{\mu\nu} \partial_\mu \phi \partial_\nu \phi \right). \tag{1}$$

Symmetry breaking fixes the vacuum expectation value of $\phi$, so $\phi^2$ attains a positive value $\langle \phi \rangle^2$, and to obtain the standard Einstein Lagrangian in the broken symmetry regime we need to select the $+$ sign in the expression (1). This inflicts on our theory a kinetic term of the wrong sign, which implies a ghost and a disastrous instability of the $\phi$ field in the pre-symmetry breaking regime. For the Feynman propagator of the $\phi$ field, the wrong sign of the kinetic term implies a wrong sign for the residue at the pole, and a violation of unitarity. This notorious sign problem of the kinetic term afflicts a multitude of models that seek to implement conformal invariance by insertion of scalar fields in the Lagrangian [2, 7, 13, 14, 15, 16, 17].

The usual attitude is to ignore this problem, by the specious argument that the wrong sign of the kinetic term is irrelevant because the entire term is eliminated when we assign to the scalar field a fixed value. This fixing is often characterized as a mere choice of gauge. But that is a misconception, because by fixing the scalar field at a constant value we not only make a choice of gauge, but we also break the conformal symmetry, and we rob the scalar field of all its dynamics (as other authors have commented, the scalar field "ceases to be a degree of freedom altogether" and is reduced to a "gauge artifact" [9]; or we "put one degree of freedom to zero, such as to get rid of the ghost" [17]).

In essence, this way of dealing with the wrong sign of the kinetic term is an argument *against* conformal symmetry, because it says that to avoid the wrong sign we need to break and remove the conformal symmetry. And, concomitantly, by setting the value of the scalar field equal to a constant of dimension $\propto$ mass, we introduce a dimensional constant into the Lagrangian, spoiling any hope for renormalizability of the theory.

As an alternative to this gauge fixing, Jackiw and Pi [18] recently proposed removal of the scalar field by means of a change of variables from what is called the "Jordan frame" to the "Einstein frame." They designate the metric tensor in the Lagrangian (1) by $g^J_{\mu\nu}(x)$, and they introduce a new metric tensor by a change of variables $g^E{}_{\mu\nu}(x) \equiv \phi^2(x) g^J{}_{\mu\nu}(x)$ .[1] By this change of variables, they give the Jordan Lagrangian (1) the form of the standard Einstein Lagrangian $\sqrt{-g^E} \frac{1}{2} R(g^E{}_{\mu\nu})$ . They then claim that this Einstein Lagrangian "clearly lacks local conformal symmetry." But this claim is fallacious: under conformal transformations, $g^J{}_{\mu\nu} \to e^{2\alpha} g^J{}_{\mu\nu}$ and $\phi \to e^{-\alpha}\phi$ , and therefore $g^E{}_{\mu\nu} = \phi^2 g^J{}_{\mu\nu} \to (e^{-\alpha}\phi)^2 (e^{2\alpha} g^J{}_{\mu\nu}) = g^E{}_{\mu\nu}$ , which means $g^E{}_{\mu\nu}$ is invariant under conformal transformations, and so is the Einstein Lagrangian constructed with $g^E{}_{\mu\nu}$ . The same is true for all other terms that might occur in a general Lagrangian endowed with conformal invariance, including terms that involve fermions, bosons, and their gauge fields—the change of variables from Jordan to Einstein changes the form of these terms, but not their invariance. This persistence of conformal invariance is no surprise, because a mere change of variables cannot change the overall invariance properties of a given Lagrangian, if the symmetry transformations of the changed variables are properly taken into account.

Jackiw and Pi's characterization of the conformal invariance as a "fake gauge invariance," contrived by "introducing a spurion field and dressing up a model to appear gauge invariant," is a misconception arising from their failure to recognize that the conformal invariance survives, even when the change of variables hides the scalar field within another field. It may be fair to characterize the scalar field as spurious, because it is merely a marker, or an "order parameter," that reveals how the conformal symmetries in the Jordan and Einstein versions of the Lagrangian are related, but the underlying symmetry does not depend on this marker—and it is not a fake symmetry. For a clearer appreciation of the meaning of conformal symmetry, it is instructive to compare the conformally symmetric Lagrangian (1) with an example of a *nonsymmetric* Lagrangian consisting of (1) plus a mass term $-\sqrt{-g} m^2 \phi^2$ . Such a mass term is not invariant under the conformal transformation, and furthermore the scalar field in this term cannot be hidden by a change of variables from Jordan to Einstein.

In the Jordan frame used for Eq. (1), resolution of the ghost problem by choice of gauge is incompatible with preservation of explicit conformal symmetry. If we want to exploit conformal symmetry for the purposes of quantum field theory, we have to ensure that this symmetry is explicit at high temperatures or high energies, and that it is not infested with fundamental inconsistencies, such as unacceptable ghosts. This makes it imperative to repair the wrong sign of the kinetic term and thereby exorcise the ghost. Furthermore, if we can repair this wrong sign and endow the field with viable, consistent dynamics, then we can attempt to add suitable interactions of the scalar field with itself and with other fields, so that the scalar field will adopt a fixed value at low energies, by the Brout-Englert-Higgs mechanism of spontaneous symmetry breaking [19, 20] .

The model presented in Sections 2 and 3 shows how this exorcism can be achieved by introducing a Weyl gauge-vector $\varphi_\mu(x)$ in conjunction with the dilaton scalar. The Weyl vector not only solves the problem of the wrong sign of the kinetic term and gives us an explicit dynamic mechanism for conformal symmetry breaking, but it also plays a crucial role in the structure of the spacetime geometry in the symmetric regime.

At present, the geometry of our universe is Riemannian, but before conformal symmetry breaking, our geometry was a Weyl geometry, devoid of well-defined absolute proper-time intervals. The geodesics in such a geometry are determined by the affine structure, that is, parallel transport of vectors tangential to the geodesic, and not by a condition of extremal proper time. According to Ehlers, Pirani, and Schild [21], fundamental axioms and propositions of differential geometry demand the existence of a Weyl vector for construction of the affine connection of this geometry, because this vector ensures that—despite the gauge dependence of the metric tensor—the geodesics are conformally invariant, as they must be, on physical grounds. After conformal symmetry breaking, the Weyl vector vanishes, and this gives us a Riemannian geometry, with a well-defined metric tensor and absolute proper-time intervals, and with geodesics determined by extremal proper time. Concurrently, some of the precursor massless fundamental particles acquire masses, which gives us the physical standards of length and time needed for the measurement of these proper-time intervals. Thus, conformal symmetry breaking leads to a wealth of new physics—and we cannot pretend that this symmetry breaking is merely a cavalier adoption of one conformal gauge choice over another, by fiat. The conformal gauge choice that emerges from the dynamics of spontaneous symmetry breaking is a symptom of drastic changes in the physics and geometry of spacetime.

---

[1] The preservation of the dimension of the metric tensor requires a dimensional factor $1/(\text{mass})^2$ on the right side of this transformation equation, to compensate the dimension of $\phi^2$ . This dimensional factor is of crucial importance in investigations of renormalizability. For the sake of simplicity, I here imitate Jackiw and Pi in omitting this factor.

My model also adopts the additional assumption that the dilaton scalar is a *complex* field $\chi(x)$. This avoids the singular behavior of the conformal Brans-Dicke equation for a real scalar field $\phi$, where the parameter value $\omega = -3/2$ appropriate for conformal symmetry collapses the field equation to a condition of zero trace for the energy-momentum tensor, while leaving the field completely undetermined [22]. In contrast, with a complex scalar field $\chi$, the field equation leaves the magnitude $\chi\chi^*$ undetermined but fully determines the evolution of the phase of the field. The phase of the scalar field is conformally invariant, whereas the freedom of choice of the magnitude of the field represents the conformal gauge symmetry. This sharp separation of the scalar field into a gauge component and a conformally invariant component makes the physics of the scalar field more transparent and reveals the close analogy between the behavior of this gauge field and other known multi-component gauge fields.

Furthermore, the adoption of a complex scalar $\chi$ permits my model to proceed by direct imitation of the Coleman-Weinberg model for massless scalar "electrodynamics" [23]. As in the latter model, the Weyl vector partially absorbs the complex scalar $\chi$ by the usual Brout-Englert-Higgs mechanism but leaves a residual real massive scalar field. This residual scalar field might serve as a significant WIMP contribution to the missing dark mass in and around galaxies, which might provide direct observational evidence for or against the model.

Another advantage of a complex scalar is that the current density $\chi\partial_\mu\chi^* + \chi^*\partial_\mu\chi$ for the wave field of individual $\chi$ particles vanishes, so any primordial particles of this kind, left over from an early stage of the universe, are incapable of acting as sources of the Weyl gauge vector to which this current couples [see Eq. (13)]. In contrast, for a real scalar field, the current would be nonzero, which could lead to an undesirable cosmological Weyl vector, in conflict with the required vanishing of the Weyl vector in the Riemannian geometry that emerges from symmetry breaking.

In my proposed Lagrangian the correction of the notorious wrong sign of the scalar kinetic term is achieved by adopting a hybrid combination of the usual conformally invariant expression with a kinetic term of the wrong sign,

$$\sqrt{-g}\,\sinh^2\Theta(\tfrac{1}{6}\chi\chi^*R - g^{\mu\nu}\partial_\mu\chi\,\partial_\nu\chi^*),\qquad(2)$$

and a new expression with a kinetic term of the right sign,

$$\sqrt{-g}\,\cosh^2\Theta\,g^{\mu\nu}(\partial_\mu - \tfrac{1}{2}b\varphi_\mu)\chi(\partial_\nu - \tfrac{1}{2}b\varphi_\nu)\chi^*.\qquad(3)$$

Here the parameter $\Theta$ is a "mixing angle" which ensures that the combination of the kinetic terms in the expressions (2) and (3) yields the correct net value $\sqrt{-g}\,g^{\mu\nu}\partial_\mu\chi\,\partial_\nu\chi^*$ with the correct sign, so there is no ghostery.

In the expression (3), $\partial_\mu - \tfrac{1}{2}b\varphi_\mu$ is a conformally gauge covariant derivative operator analogous to the familiar gauge covariant derivative operator $\partial_\mu + ieA_\mu$ of electrodynamics. The magnitude of the dimensionless coupling constant $\tfrac{1}{2}b$ that multiplies the Weyl vector field $\varphi_\mu$ (analogous to the dimensionless electric coupling constant $e \cong 1/\sqrt{137}$ that multiplies the vector field $A_\mu$ of electrodynamics) is set by the geometric interpretation of $\varphi_\mu$ as the gauge vector for the transport of lengths and time intervals (see Section 4), as originally proposed by Weyl [24, 25]. For the Coleman-Weinberg calculation in Section 3, the factor $\cosh^2 \Theta$ must be included in the coupling constant. In principle, any nonzero value of $\Theta$ is permissible, but in practice it will be convenient to assume that $(\frac{1}{2} b \cosh^2 \Theta)^2$ is small, so that perturbation theory is applicable.

The kinetic terms of other scalar fields—such as, say, the Higgs field—are assumed to appear in the Lagrangian only in the conformally invariant combination $\sqrt{-g}\,[-\frac{1}{6} H H^\dagger R + g^{\mu\nu} D_\mu H (D_\nu H)^\dagger]$, with the correct positive sign for the kinetic term and the negative sign shifted to the $R$ term (the differential operator $D_\mu$ is the appropriate gauge covariant derivative constructed with the $\mathrm{SU}(2) \times \mathrm{U}(1)$ gauge fields of the Standard Model, but it does not include the Weyl gauge field $\varphi_\mu$). The correct final positive sign for the net $R$ term, after all symmetry breakings have been completed, is achieved by assuming that the positive-sign contribution arising from the vacuum expectation value of $\chi\chi^*$ is larger than the sum of negative-sign contributions arising from the vacuum expectation values of $H H^\dagger$ and other scalar fields implicated in symmetry breakings.

Conformally covariant derivative operators with a Weyl vector were used in Lagrangian models by Smolin [26], Cheng [27], Nishino and Rajpoot [28], and Drechsler and Tann [29]. My model is closest to that of Smolin, who, however, uses a real scalar field. As mentioned above, a real scalar field has various disadvantages in comparison with a complex field. Furthermore, spontaneous symmetry breaking based on a single-component real scalar field is problematic. The usual Englert-Brout-Higgs mechanism for a scalar interacting with a massless vector field requires a complex scalar field, that is, two real scalars, so that one of these scalars can be "gauged away" and "eaten up" by the vector field, which thereby acquires a mass. Smolin concedes it is "unlikely that a single-component [real] scalar field will in fact develop an expectation value as a result of quantum corrections." This problem does not arise when the scalar field is complex, because then the explicit scalar-vector interactions in the Lagrangian (4) directly lead to one-loop quantum effects that generate an effective potential for the vacuum expectation value of the scalar field, in imitation of Coleman-Weinberg massless scalar electrodynamics. Instead of adopting this direct and explicit approach, Smolin engages in vague speculations about unspecified instabilities in the full quantum theory that might contribute to his Lagrangian an extra, unspecified, effective potential, which might favor an asymmetric vacuum for the scalar field.

The other Lagrangian models with Weyl vectors [27, 28, 29] suffer from the same troubles as that of Smolin, with an extra defect in that the authors inserted gauge-covariant derivative operators $\partial_\mu - \frac{1}{2} b \varphi_\mu$ somewhat indiscriminately into every which derivatives in their Lagrangians, including the derivatives of fermion fields and boson fields. This leads to unacceptable consequences, because if the fermion fields act as sources for the Weyl vector field, then the 0 component of the Weyl vector field would persist after conformal symmetry breaking, with a value of approximately $\varphi_0 \sim$ (fermion density)$/m_P^2$, so the high fermion density of the early universe would maintain a large value of $\varphi_0$ until long after conformal symmetry breaking. This would give rise to a perplexing conflict between the conformally non-symmetric and unambiguous Riemannian geometry and a concurrent Weyl transport law for length and time intervals—the geometry would be neither fish nor fowl, partly a Riemannian geometry and partly a Weyl conformal geometry, with abnormalities in the transport behavior of clock rates and atomic frequencies, such as the predicted abnormalities that originally led Einstein and others to reject Weyl's theory [30]. My model resolves this dilemma by adopting the scalar field $\chi$ as the one and only source for the Weyl vector field, so when the components of the current density of this scalar field vanish upon conformal symmetry breaking, the Weyl vector field also vanishes (at least as a classical field, although a gas of incoherent Weyl quanta might survive and perhaps make a significant contribution to the clouds of dark mass in and around galaxies).

Accordingly, in my proposed Lagrangian the conformally covariant derivatives appear only in connection with the scalar field $\chi$.

My paper and that of Bars et al. should be regarded as complementary. For instance, for the Bezrukov-Shaposhnikov model, the Bars et al. "conformalization" prescription yields a conformally-invariant precursor Lagrangian for the Higgs field, while my hybridization procedure eliminates the unacceptable ghost in the dilaton field used for this conformalization. A fortunate feature of the Bars et al. conformalization is that it relies entirely on the dilaton scalar, and does not involve the Weyl vector at all, so the Higgs field does not become an undesirable source of Weyl vector fields after conformal symmetry breaking.

Sections 2 and 3 of my paper show how the hybridization of dilaton kinetic terms not only exorcises the ghost, but also provides an explicit mechanism for conformal symmetry breaking, in almost exact imitation of Coleman-Weinberg model of massless scalar electrodynamics. The masses for the dilaton scalar and the Weyl gauge-vector that emerge from this model by spontaneous symmetry breaking via the Brout-Englert-Higgs mechanism are of the order of 1/10 of the Planck mass or somewhat smaller. Accordingly, the conformal symmetry breaking occurs somewhat later than the Planck time, which is consistent with the view that this symmetry breaking is not directly related to quantum effects of the geometry.

Section 4 considers the geometrical interpretation of the Weyl gauge-vector as a device for the transport of length and time intervals, transport of parallels, construction of the affine connection and geodesics, and construction of a "proper" metric tensor which defines conformally invariant lengths along individual non-null worldlines, but only in a path-dependent manner.

## 2 The conformal model

The arguments presented in the preceding section motivate my proposal for a simple model for breaking gravitational conformal symmetry with the following assumptions (essentially the same as those already stated in a preliminary version of this paper [31]): (*i*) Einstein's theory should emerge from this symmetry breaking, except for some macroscopically undetectable corrections near the Planck length; (*ii*) before symmetry breaking, the model consists of a conformally-invariant version of the Jordan [32] and Brans-Dicke [22] theories (with $\omega = -3/2$ and $T_\mu^\mu = 0$), but with a complex scalar field $\chi(x)$, which is coupled not only to gravitation but also to a Weyl gauge-vector field $\varphi^\mu$ by a "conformal current" $\partial \mathcal{L} / \partial \varphi_\mu$ carried by the scalar field, so this current acts as source of the vector field; (*iii*) exactly as in the Coleman-Weinberg model for massless scalar electrodynamics [23], an effective potential for the vacuum expectation value of the scalar field arises from radiative corrections to the $\chi$-$\varphi_\mu$ interaction, supplemented by a conformally-invariant self-interaction $(\chi\chi^*)^2$; (*iv*) spontaneous symmetry breaking then occurs by the Brout-Englert-Higgs mechanism, resulting in large masses for the residual (real) scalar and the vector $\varphi_\mu$.

The proposed conformally-invariant Lagrangian density is (with the signature $+---$)

$$\mathcal{L} = \sqrt{-g} \sinh^2 \Theta \left[ \tfrac{1}{6} \chi\chi^* R - g^{\mu\nu} \partial_\mu \chi \, \partial_\nu \chi^* \right] + \sqrt{-g} \cosh^2 \Theta \left[ g^{\mu\nu} (\partial_\mu - \tfrac{1}{2} b\varphi_\mu) \chi \, (\partial_\nu - \tfrac{1}{2} b\varphi_\nu) \chi^* \right.$$
$$\left. - \tfrac{\lambda}{4!} (\chi\chi^*)^2 \right] - \tfrac{1}{4} \sqrt{-g} \, g^{\mu\sigma} g^{\nu\tau} f_{\mu\nu} f_{\sigma\tau} + \mathcal{L}_{(m)}, \tag{4}$$

where $f_{\mu\nu} \equiv \partial_\nu \varphi_\mu - \partial_\mu \varphi_\nu$, and where $\mathcal{L}_{(m)}$ is the conformally-invariant Lagrangian density for the various fermion, boson, and gauge fields associated with the strong and electroweak interactions and their characteristic scalar fields (such as the Higgs field), before breaking of their own gauge symmetries. In the Lagrangian (4) all these fundamental matter fields are assumed to be massless, and the observed masses of the known particles are assumed to arise at a later stage by spontaneous symmetry breaking of their own gauge symmetries, as in the case of the Higgs boson. (For fermions, the contribution to $\mathcal{L}_{(m)}$ must be expressed in terms of tetrad vectors $V_\mu^{(\sigma)}(x)$ instead of the metric tensor, which becomes $g_{\mu\nu} = V_\mu^{(\sigma)} V_\nu^{(\tau)} \eta_{\sigma\tau}$. This makes the fermion Lagrangian rather messy; but the conformal gauge transformation of a tetrad vector is simple, merely multiplication by a factor $\mathrm{e}^{\alpha(x)}$, where $\alpha(x)$ is the same arbitrary function used in Eq. (5). In the following calculations, no explicit use will be made of these details for fermions, which are readily available in textbooks [33]).

The Lagrangian density (4) is invariant under the conformal gauge transformations

$$g_{\mu\nu}(x) \to g_{\mu\nu}(x)\mathrm{e}^{2\alpha(x)} ,$$
$$\chi(x) \to \mathrm{e}^{-\alpha(x)}\chi(x) ,$$
$$\varphi_\mu(x) \to \varphi_\mu(x) - (2/b)\partial_\mu\alpha(x) . \tag{5}$$

where $\alpha(x)$ is an arbitrary real function. Note that the conformal transformations generically involve a curved spacetime geometry, and that these transformations defined by Eq. (5) must not be confused with the so-called scale transformations in a flat spacetime with Cartesian coordinates, for which the transformation of, say, a scalar field has a different definition, $\chi(x) \to \mathrm{e}^{-\alpha}\chi(\mathrm{e}^{-\alpha}x)$ where $\alpha$ is a *constant* real number [34]. Such a scale transformation can be regarded as a compound transformation consisting of a constant conformal transformation of flat spacetime followed by a coordinate transformation $x \to \mathrm{e}^{-\alpha}x$; thus, for a Lagrangian expressed in general coordinates, scale invariance is implicit in conformal invariance, although only in flat spacetime. Symmetry breaking of scale-invariant Lagrangians has recently been explored as a possible mechanism for generating the Higgs mass [35], but this is a very questionable approach because such scale transformations are applicable only in flat spacetime and cannot be generalized to curved spacetime.

The various fields contained in $\mathcal{L}_{(m)}$ have their own appropriate conformal transformations, but we will not need the full details of these transformations in the following calculations. Conformal invariance for scalar fields in $\mathcal{L}_{(m)}$ other than $\chi$ is to be achieved by combining each kinetic term with a term proportional to $R$ (with a negative sign) as in the conformally invariant combination $\sqrt{-g}\,[-\frac{1}{6}HH^\dagger R + g^{\mu\nu}D_\mu H(D_\nu H)^\dagger]$ for the Higgs field already mentioned in Section 1, and by applying the conformalization procedure of Bars et al. to any scalar interaction terms, as appropriate. Conformal invariance of contributions from massless fermion and boson fields (other than scalar fields) is automatic—it requires no modifications of the general coordinate invariant kinetic terms of these fields. Accordingly, $\mathcal{L}_{(m)}$ does not contain the field $\varphi_\mu$ at all and can therefore be ignored in the following discussion of the dynamics of conformal symmetry breaking caused by the $\chi$ - $\varphi_\mu$ interaction.

It is possible to add to the Lagrangian (4) a conformally-invariant term containing products of second-order derivatives of $g_{\mu\nu}$, leading to fourth-order derivatives in the field equations. But in my model such extra terms play no direct role in achieving symmetry breaking, so I will ignore them for now. [2] This is in contrast to some other attempts at generating ordinary gravity from a conformally-invariant Lagrangian, in which a combination of quadratic products of second-order derivatives (the square of the conformal Weyl tensor) plays an essential role, but no term proportional to $R$ is included in the Lagrangian, at least not ab initio [4, 5, 8, 10].

---

[2] With the Lagrangian (4), additional conformally-invariant higher-order derivative terms, such as the term $\alpha_{grav}\sqrt{-g}\,(R_{\mu\nu}R^{\mu\nu} - R^2/3)$ often favored by theorists, merely add short-range Yukawa potentials to the usual macroscopic $1/r$ Newtonian potential. If $\alpha_{grav}$ is of the order of magnitude of $\sim 1$, then the range of this Yukawa potential is about a Planck length and it produces no measurable macroscopic effects. However, the higher-order derivatives can lead to drastic modifications of the singularities found in general relativity.

## 3 Results

In my model, the current density of the scalar field [see Eq. (13)] and its coupling to the vector field are not the same as in the Coleman-Weinberg scalar electrodynamics model [23]. However, the coupling $\chi\chi^*\varphi_\mu\varphi^\mu$ is exactly the same, with the substitution of $(b/2)^2\cosh^4\Theta$ for $e^2$. The one-loop calculation of the effective potential does not depend on the current, but only on the $\chi\chi^*\varphi_\mu\varphi^\mu$ coupling and on the quartic self-coupling $(\lambda/4!)\cosh^2\Theta(\chi\chi^*)^2$. Equation (4) therefore leads to an effective potential for the vacuum expectation value $|\chi_c|$ of the same form of that of Coleman-Weinberg, with only some trivial changes in the coupling constants:

$$V_{eff}(|\chi_c|) = \frac{3b^4\cosh^8\Theta}{(32\pi)^2}|\chi_c|^4\left(\ln\frac{|\chi_c|^2}{\langle\chi\rangle^2}-\frac{1}{2}\right), \tag{6}$$

where $\langle\chi\rangle$ is the value of $|\chi_c|$ at which the minimum occurs. Equation (6) includes renormalization of a divergent integral and also includes the Coleman-Weinberg "dimensional transmutation," by means of which the dimensionless coupling constant $\lambda$ is "traded" for a dimensional constant $\langle\chi\rangle$. When the field settles into the stable minimum value $\langle\chi\rangle$, the Higgs mechanism breaks the conformal symmetry and endows the scalar and vector fields with masses. With $|\chi_c|$ fixed at $\langle\chi\rangle$, the first term in the Lagrangian (4) then becomes $\frac{1}{6}\langle\chi\rangle^2\sinh^2\Theta\sqrt{-g}R$. For agreement with Einstein's theory in the classical regime, we make the choice

$$\frac{1}{6}\langle\chi\rangle^2\sinh^2\Theta = \frac{m_P^2}{16\pi}. \tag{7}$$

The resulting vector and scalar masses are then, respectively,[3]

$$m_V^2 = \frac{3b^2}{32\pi}\frac{\cosh^4\Theta}{\sinh^2\Theta}m_P^2 \tag{8}$$

and

$$m_S^2 = \frac{1}{\pi}\left(\frac{3b^2}{32\pi}\right)^2\frac{\cosh^8\Theta}{\sinh^2\Theta}m_P^2. \tag{9}$$

---

[3] The masses are here expressed in terms of $\Theta$, but they can be alternatively expressed in terms of $\lambda$ because the condition for a minimum in the effective potential implies the relation $\lambda = (33/128\pi^2)b^4\cosh^6\Theta$ between the coupling constants [23].

If the angle factors are of the order of ~ 1, then $m_V$ is of the order of ~ $bm_P/10$ and $m_S$ is ~ $b^2 m_P/100$, so gravitational and geometric effects of the scalar and vector fields are not accessible to macroscopic measurements.[4] After symmetry breaking, neither the scalar field nor the vector field reveal themselves at the macroscopic level, and we can ignore the effects of the Weyl gauge-vector on the transport of lengths (see Section 4) and the corresponding modifications of Riemannian differential geometry. This is in contrast to the standard Brans-Dicke theory, in which the massless scalar field makes a contribution to long-range gravitational effects, and thereby causes measurable violations of the equality of inertial and gravitational masses for systems of particles or fields containing gravitational self-energy. Before symmetry breaking, the zero-mass, long-range, scalar and vector fields in the Lagrangian (4) also make such abnormal contributions to the gravitational masses; but this does not affect the macroscopic free-fall experiments that we perform today.

The depth of the effective potential (6) at its minimum is about $-b^4 m_P{}^4/10^4$, which indicates that the spontaneous breaking of conformal symmetry in the early universe occurs at a characteristic thermal energy $kT \approx bm_P/10$. For our universe, with a Friedmann-Lemaître geometry, this corresponds to a time of about $10/bm_p$. If the coupling constant is reasonably small, say, $b \approx 1/100$ or $1/1000$, then this is significantly later than the Planck time, which suggests that quantum gravity does not play a major role in the symmetry breaking.

## 4  The Weyl gauge-vector and length transport

Various expositions of Weyl geometry are readily available [24, 25, 30, 36, 37, 38, 39, 40], of which the clearest is that by Dirac [37], although it is somewhat blighted by a stubborn insistence on following Weyl's misbegotten notions about the geometrization of electrodynamics. A brief review of Weyl geometry therefore seems worthwhile, with emphasis on a few points that have not received the attention they deserve.

Before symmetry breaking, the conformal gauge transformation $g_{\mu\nu} \rightarrow g_{\mu\nu} e^{2\alpha(x)}$ of the metric tensor prevents us from associating an unambiguous, absolute, length or time with small displacements $\Delta x^\kappa$ at different locations. At a fixed location, we can compare lengths squared $\ell^2 = g_{\kappa\lambda} \Delta x^\kappa \Delta x^\lambda$ in different directions, but we cannot compare lengths at different locations because such a comparison is gauge-dependent. Weyl argued that to achieve comparisons of lengths at different locations it is necessary to generalize Riemannian differential geometry by a supplementary transport law for lengths that incorporates a gauge-vector field $\varphi_\mu$, so the change in a small length squared $\ell^2 = g_{\kappa\lambda} \Delta x^\kappa \Delta x^\lambda$ subjected to a transport $dx^\mu$ is [25][5]

$$d(\ell^2) = -b\ell^2 \varphi_\mu dx^\mu. \qquad (10)$$

In an incisive general analysis of non-Riemannian geometries, Ehlers et al. [21] showed that this transport law for lengths is actually a consequence of the affine structure of Weyl's geometry, and that the Weyl gauge vector $\varphi_\mu$ is a required feature of this geometry, according to fundamental theorems of differential geometry based on reasonable, intuitively self-evident, axioms about light rays and particle worldlines. The Weyl geometry is endowed with a "conformal" structure consisting of well-defined light cones and time-like, space-like, and null directions, in conjunction with a "projective" structure consisting of geodesic worldlines representing the motions of freely-falling particles of nonzero mass. (Conformal invariance of the Lagrangian demands that the fundamental particles in the Lagrangian be massless, so it would seem that only null-geodesics have physical significance when the universe is in a conformally invariant regime. However, this does not forbid us to contemplate hypothetical non-null geodesics, in the same way we might contemplate hypothetical spacelike geodesics in, say, Minkowski spacetime.)

---

[4] For conformal invariance, each of the Higgs scalar fields $H$ associated with the breaking of GUT and electroweak symmetries requires the addition of an extra term $-\sqrt{-g}HH^\dagger R/6$ to the Lagrangian. This alters Eq. (7) and requires a small increase of $\langle \chi \rangle$, which leads to small changes in the masses $m_V$ and $m_S$.

[5] Weyl did not include the adjustable coupling constant $b$ in his law, and he used the symbol $\ell$ for the length squared of a vector, whereas I prefer $\ell^2$.

The joint conformal and projective structures imply an affine structure with a parallel-transport law, and according to Ehlers et al. that is where the Weyl vector makes its debut [see Eq. (12)]. The Weyl transport law for lengths stated in Eq. (10) is then a consequence of the parallel-transport law, rather than viceversa. However, the arguments of Ehlers et al. are rather intricate, and for the sake of brevity I will here accept Weyl's transport law for lengths as a starting point and proceed from this to the parallel-transport law.

For a parallel-transport law of the usual form, $d(\Delta x^\kappa) = -\Gamma^\kappa{}_{\nu\mu}\Delta x^\nu dx^\mu$, the transport of the displacements contained in $\ell^2$ on the left side of Eq. (10) leads to

$$d(\ell^2) = \frac{\partial g_{\kappa\lambda}}{\partial x^\mu} dx^\mu \Delta x^\kappa \Delta x^\lambda - g_{\kappa\lambda}\Gamma^\kappa{}_{\nu\mu}\Delta x^\nu \Delta x^\lambda dx^\mu - g_{\kappa\lambda}\Gamma^\lambda{}_{\nu\mu}\Delta x^\kappa \Delta x^\nu dx^\mu. \qquad (11)$$

The requirement that this equal the right side of Eq. (10) then gives us a set of equations that we can solve for the affine connection coefficients:

$$\Gamma^\alpha{}_{\mu\nu} \equiv \tfrac{1}{2} g^{\alpha\beta}(g_{\beta\mu,\nu} + g_{\nu\beta,\mu} - g_{\mu\nu,\beta}) \;+ \tfrac{1}{2} b(\delta^\alpha_\mu \varphi_\nu + \delta^\alpha_\nu \varphi_\mu - g_{\mu\nu}\varphi^\alpha). \qquad (12)$$

These $\Gamma$ coefficients must be gauge invariant, because the parallel transport of vectors cannot depend on the choice of gauge. With the gauge transformation $g_{\mu\nu} \to e^{2\alpha} g_{\mu\nu}$ for the metric tensor, Eq. (12) then requires that $\varphi_\mu \to \varphi_\mu - (2/b)\partial_\mu \alpha(x)$. This confirms the consistency of the transport law (10) with the field-theoretic treatment of the gauge vector in Section 2.

In view of the similarity between the gauge transformation of $\varphi_\mu$ and that of the electromagnetic vector potential $A_\mu$, Weyl rashly proposed that $\varphi_\mu$ should be identified with $A_\mu$, and he thought he could thereby achieve a geometrical interpretation of the electromagnetic field. But if we attempt to make this identification in the Lagrangian (4), we find that the current density that acts as source for the gauge-field $\varphi^\mu$ is the "conformal current"

$$\frac{\partial \mathcal{L}}{\partial \varphi_\mu} = -\tfrac{1}{2} b\sqrt{-g}\cosh^2\Theta[\chi(\partial^\mu - \tfrac{1}{2}b\varphi^\mu)\chi^* + \chi^*(\partial^\mu - \tfrac{1}{2}b\varphi^\mu)\chi], \qquad (13)$$

which disagrees with the usual electric current. This disagreement precludes the naïve identification of Weyl's gauge-vector with the electromagnetic vector potential.

The conformal current density given by Eq. (13) is conserved, as can be seen directly from the field equation $\partial_\nu(\sqrt{-g}\,f^{\mu\nu}) = \partial \mathcal{L}/\partial \varphi_\mu$. Alternatively, this conservation law for the current can be derived from invariance of the Lagrangian under conformal transformations, by Noether's theorem. The expression on the right side of Eq. (13) is the Noether current that arises from the $\cosh^2\Theta$ term of the Lagrangian (4). This is the only term that contributes a current, because all the other terms in my Lagrangian can be entirely expressed as functions of conformally invariant variables, such as $g^E{}_{\mu\nu}$, and this makes it immediately obvious that these terms do not contribute anything at all to the Noether current, so Eq. (13) is actually the total current. In their critique of conformal symmetry, Jackiw and Pi [18] regarded the vanishing of the Noether current in their model (which lacks the $\cosh^2\Theta$ term) as evidence that the conformal symmetry plays no dynamical role. My model avoids this pitfall, because, in the conformally-symmetric regime, the $\cosh^2\Theta$ term in Eq. (13) evidently does not vanish.

In the symmetry-broken regime, with $\chi\chi^* = \text{constant}$, the derivative terms on the right side of Eq. (13) vanish, and the nonderivative terms are of the form $\varphi^\mu \times \text{constant}$, which is merely a mass term already included in the mass formula (8) of the Coleman-Weinberg calculation. Thus, the gauge field $\varphi^\mu$ effectively becomes sourceless and ceases to exist as a static or quasistatic classical field. This is consistent with the geometric requirement that in the symmetry-broken regime the Weyl vector must be zero, so lengths are transported in the normal way expected in a metric Riemannian geometry.

The familiar Christoffel symbols in the first line of Eq. (12) completely determine the null geodesics. The combination of Weyl-vector derivatives in the second line of Eq. (12) has no effect on null geodesics—it merely alters the null geodesic by a reparametrization. This insensitivity of null geodesics to the Weyl vector is consistent with the absence of the Weyl vector from the conformally-invariant Lagrangian $\mathcal{L}_{(m)}$ for massless fields, whose field equations have free-wave solutions that propagate along light cones. But for non-null geodesics, the combination Weyl-vector derivatives in the second line of Eq. (12) plays a crucial role, and it cannot be ignored.

Note that the $\Gamma$ coefficients in Eq. (12) are symmetric in their lower indices $\mu$-$\nu$. This means that the Weyl geometry has no torsion. Proposed gravitational theories with torsion, such as the Einstein-Cartan-Kibble theory [41], attempt to play a game of hide-and-seek by making the torsion directly proportional to the local spin density, so it exists only at the location of particles and vanishes in the empty regions of space between and around particles. But it is not clear whether this trick can save such theories from a conflict with the fundamental requirements of a Weyl conformal geometry.

By integration of the transport law (10) along a specified worldline we can determine an absolute length for small intervals $\Delta x^\kappa$ along or near this worldline, relative to a standard of length established at a given initial reference point or else relative to a standard of length established on a given reference hypersurface which all the contemplated worldlines intersect and on which conformal symmetry is broken, so there exist absolute, calibrated, standards of length based on particle masses (for instance, for purposes of cosmology, we might adopt the equal-time hypersurface of our universe today as our reference surface and integrate backward in time along a worldline into the conformal era of the early universe). Integration of the transport law (10) tells us that the ratio of small lengths squared at the reference point $x = 0$ and at an arbitrary point on the specified worldline is

$$\frac{\ell^2(x)}{\ell^2(0)} = \exp\left[-b\int_0^x \varphi_\mu\, dx^\mu\right]. \tag{14}$$

Accordingly, we can define a recalibrated metric tensor $\tilde{g}_{\mu\nu}$ at any point on this worldline,

$$\tilde{g}_{\mu\nu}(x) \equiv \frac{\ell^2(0)}{\ell^2(x)} g_{\mu\nu}(x) = g_{\mu\nu}(x)\exp\left[b\int_0^x \varphi_\mu\, dx^\mu\right]. \tag{15}$$

This recalibrated metric tensor, or *proper* metric tensor, has no global significance—it is a functional of the worldline, and it is valid only along the selected worldline (when two worldlines intersect, they will usually have different proper metric tensors at the intersection point).

Under the gauge transformations for $g_{\mu\nu}$ and $\varphi_\mu$, with $\alpha(0) = 0$, the proper metric tensor $\tilde{g}_{\mu\nu}(x)$ is conformally invariant, and it assigns a conformally-invariant, absolute, length to intervals along the given worldline. Thus, the proper time measured along the worldline is defined unambiguously, even though the background geometry is conformally symmetric. We therefore find that the geodesics defined by the affine structure of the Weyl geometry are equivalent to the geodesics defined as worldlines of extremal length calculated according to their proper metric tensor. It's a case of "metric geometry without metric geometry," but only along each selected worldline, with a suitable initial reference point $x = 0$.

This pseudo-metric behavior, subject to the restriction of selected worldlines, also applies to the affine connection coefficients given by Eq. (12). Along a worldline, with a proper metric tensor $\tilde{g}_{\mu\nu}$ defined according to Eq. (15), the affine connection $\Gamma^{\alpha}{}_{\mu\nu}$ can be shown to be identical to the usual metric affine connection (that is, the Christoffel symbol) calculated from $\tilde{g}_{\mu\nu}$:

$$\Gamma^{\alpha}{}_{\mu\nu} = \tilde{\Gamma}^{\alpha}{}_{\mu\nu} = \tfrac{1}{2}\tilde{g}^{\alpha\beta}(\tilde{g}_{\beta\mu,\nu} + \tilde{g}_{\nu\beta,\mu} - \tilde{g}_{\mu\nu,\beta}), \tag{16}$$

where it is assumed that the path-dependent exponential function in Eq. (15) is differentiated by the rule that the path is held fixed and that the extra displacement $dx^{\sigma}$ used in for differentiation is added at the end of the path. The concordance of Eqs. (12) and (16) establishes that the geodesic determined from the proper tensor $\tilde{g}_{\mu\nu}$ is both a worldline of extremal length and a worldline generated by parallel transport of its tangential vector, as in metric Riemannian geometry.

Because of their geometrical significance and gauge invariance, it seems obvious that the geodesics of the proper metric tensor should be used for investigations of geodesic completeness, and that no gauge condition need be imposed on the scalar $\chi$ and the Weyl vector $\varphi_{\mu}$. Instead, the scalar and the Weyl vector generated by this scalar should be calculated from their field equations before proceeding to the determination of the geodesics. Of course, while solving the field equations we might find it expedient to adopt some specific choice of gauge, which will do no harm, because the geodesic equation to be solved subsequently is gauge invariant.

In their investigations of geodesics, Bars et al. [9] indeed adopted a specific gauge, and they took care to make their extremum principle for their geodesic equation gauge invariant. However, although gauge invariance is a necessary requirement for the geodesic equation, it is not a sufficient requirement, and the geodesic equation of Bars et al. is suspect because it contains no Weyl vector. According to the precepts of Ehlers et al., we should obtain the geodesics from the affine structure of the geometry, that is, from parallel transport; and the Weyl vector is required to convert these parallel-transport geodesics into extremal paths for proper time. Leaving the Weyl-vector out of the geodesic equation is acceptable for null geodesics, but is not acceptable for non-null geodesics. The total absence of a Weyl vector in the geodesic equation of Bars et al. would seem to indicate that one or several of the axioms stipulated by Ehlers et al. are being violated.

# 5 Conclusions

By inclusion of a Weyl gauge vector and a complex dilaton scalar with a hybrid kinetic term, the Lagrangian proposed in Section 2 eliminates the ghost problem and thereby offers a consistent model that achieves conformal invariance for the gravitational Lagrangian and also an explicit and well-founded mechanism to break this symmetry and bring the theory into agreement with conventional Einstein theory, in the low-energy regime. Conformally symmetric modifications of GUT and electroweak interactions can be included in the model by adopting the "lifting" scheme of Bars et al. for the relevant Lagrangian terms contributed by the Standard Model. The absence of dimensional constants in the Lagrangian of the resulting "full" conformally symmetric theory opens the path to a renormalizable quantum theory of gravitation. The Lagrangian (4) is not only power-counting renormalizable, but, according to a path-integral calculation by Haba [42], the singularities in the scattering matrix and the n-point functions in Brans-Dicke theory are mild, no worse than the singularities in $\phi^4$ perturbation theory, which lends support to a diagnosis of renormalizability.

Some alternative ways to formulate conformally symmetric renormalizable theories proceed without the scalar dilaton field and instead attribute the breaking of conformal symmetry to the masses that arise in the GUT or electroweak transition. The conformally invariant gravitational Lagrangians in such theories are constructed by relying on higher-order derivatives of the metric tensor, for instance, Lagrangians with quadratic products of second-order derivatives [4, 5, 8, 10]. But such higher-derivative theories require drastic and controversial modifications of standard quantum theory because of the violations of unitary with which they are afflicted [5, 13]. Furthermore, with higher-derivative gravitational field equations it is difficult or impossible to mimic the behavior of the conventional Einstein theory and obtain the standard Newtonian $1/r$ potential [43, 44, 45]. Thus, a conformally invariant  theory that merges into the conventional Einstein theory by breaking of the conformal symmetry is by far the most elegant way to accommodate the empirical facts.

Finally, it is worth emphasizing the crucial role played by the Weyl gauge vector field in conformal geometry. Attempts at conformally invariant theories have almost always proceeded without this Weyl vector and/or without due consideration of its geometrical implications. A curious result obtained by Bars et al. [9]—who claim that for non-null geodesics in the early universe, geodesic completeness/incompleteness depends on the choice of their "c-gauge" vs. their "$\gamma$ -gauge" for the dilaton scalar and the metric tensor—might well be attributable to a failure to take the Weyl vector into account. If the analysis of Ehlers et al. is correct, the absence of a Weyl vector and its geometric paraphernalia is a fatal mistake—if no Weyl vector, then no conformally invariant theory with a geometric interpretation.

## Acknowlegments

I am grateful to Hagen Kleinert, Philip Mannheim, and She-Sheng Xue for discussions of the difficulties in constructing standard gravity by symmetry breaking of a conformally-invariant theory, and I am grateful to two anonymous reviewers for bringing Refs. [18] and [35] to my attention .